\documentclass[%
reprint,
superscriptaddress,
amsmath,amssymb,
aps,
longbibliography,
floatfix,
]{revtex4-2}

\usepackage{graphicx}%
\usepackage{verbatim}
\usepackage[version=4]{mhchem}
\usepackage{xcolor}
\usepackage{physics}
\usepackage[normalem]{ulem}
\usepackage{pdfpages}
\usepackage{pgffor}
\usepackage{multirow}
\usepackage{booktabs}
\makeatletter
\AtBeginDocument{\let\LS@rot\@undefined}
\makeatother
\newcommand{\nens}{n_\text{ens}}

\newcommand{\rev}[1]{#1}

\begin{document}

\title{Uncertainty quantification by direct propagation of shallow ensembles}

\author{Matthias Kellner}
\affiliation{Laboratory of Computational Science and Modeling, Institut des Mat\'eriaux, \'Ecole Polytechnique F\'ed\'erale de Lausanne, 1015 Lausanne, Switzerland}

\author{Michele Ceriotti}
\email{michele.ceriotti@epfl.ch}
\affiliation{Laboratory of Computational Science and Modeling, Institut des Mat\'eriaux, \'Ecole Polytechnique F\'ed\'erale de Lausanne, 1015 Lausanne, Switzerland}

\date{\today}%

\begin{abstract}
Statistical learning algorithms provide a generally-applicable framework to sidestep time-consuming experiments, or accurate physics-based modeling, but they introduce a further source of error on top of the intrinsic limitations of the experimental or theoretical setup.
Uncertainty estimation is essential to quantify this error, and to make application of data-centric approaches more trustworthy. 
To ensure that uncertainty quantification is used widely, one should aim for algorithms that are accurate, but also easy to implement and apply.
In particular, including uncertainty quantification on top of an existing architecture should be straightforward, and add minimal computational overhead. 
Furthermore, it should be easy to manipulate or combine multiple machine-learning predictions, propagating uncertainty over further modeling steps. 
We compare several well-established uncertainty quantification frameworks against these requirements, and propose a practical approach, which we dub direct propagation of shallow ensembles, that provides a good compromise between ease of use and accuracy.
We present benchmarks for generic datasets, and an in-depth study of applications to the field of atomistic machine learning for chemistry and materials. These examples underscore the importance of using a formulation that allows propagating errors without making strong assumptions on the correlations between different predictions of the model.

\end{abstract}

\maketitle
\newcommand{\grayout}[1]{}

\section{Introduction}

Statistical learning frameworks and data-driven surrogate models have found broad applications in the natural sciences, physical modelling and engineering~\cite{carl+19rmp,gain+20nm,degr+22nature}. 
One particular application that we are going to focus on in this work involves the use of regression models to predict the interactions between the atomic constituents of matter (so-called machine-learned interatomic potentials (MLIPs)\cite{behl16jcp}) and more generally to approximate, at a much lower computational cost, the microscopic quantities that can be predicted by a quantum mechanical electronic-structure calculation~\cite{ceri22mrsb}.
These surrogate models enable accurate atom-scale simulations of materials of unprecedented time and length scales~\cite{weile2020pushing,deri+21nature,zhou+23ne,musaelian2023scaling}. They have been one of the early examples of application of machine learning to scientific inquiry~\cite{behl-parr07prl,bart+10prl,rupp+12prl}, and have become a well-established tool for atomic-scale modeling in chemistry and materials science~\cite{ceri+21cr}.
Besides their advantages, surrogate models add new sources of errors on top of the approximations that are inherent in the reference electronic-structure calculations. In particular, when the MLIP is applied to atomic systems that are not well represented in their training data (i.e. in the extrapolative regime) a rapid degradation of model accuracy is typically observed.
The statistical nature of ML-based surrogate model makes it possible, and desirable, to associate uncertainties with the predictions.
Uncertainty quantification (UQ) increases the level of trust in data-driven predictions, and can also be used to drive \emph{active learning} schemes~\cite{seungQueryCommittee1992, jinnouchiOntheflyMachineLearning2019a, vandermauseOntheflyActiveLearning2020a} in which large errors trigger the aquisition of new training points and retraining the model.
An important point to consider when choosing methods for uncertainty quantification is that the direct predictions of a ML model might be combined in non-trivial ways to evaluate the quantity of interest. 
This is certainly the case for  application of MLIPs in atomistic modelling, that are often used to compute  \emph{thermodynamic averages} of observables, by generating a sequence of atomic structures with a distribution that is consistent with the relevant temperature or pressure. 
This kind of tasks requires repeated model evaluation, and being able to combine uncertainty estimates to assess the error on averages.

\begin{figure*}[bthp]
\includegraphics[width=1.0\textwidth]{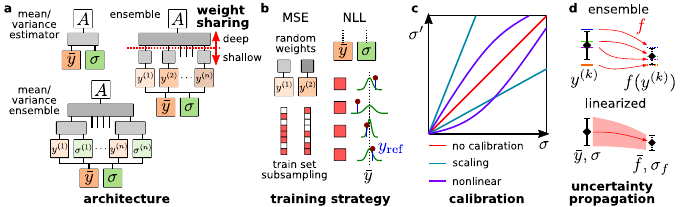}
\caption{
A schematic overview of the design space of uncertainty quantification schemes. 
Models differ  (a) by the architecture (how the uncertainty $\sigma(A)$ is evaluated); (b) by the training strategy (how the parameters are fitted to predict target and uncertainty); (c) by their calibration (whether and how the raw $\sigma(A)$ is manipulated to obtain the final prediction); (d) by the implementation of uncertainty propagation (how the uncertainty is computed when the model predictions are further manipulated to achieve the quantity of interest). 
}
\label{fig:uq-summary}
\end{figure*}

Based on these considerations, we identify four requirements for uncertainty quantification schemes, that are beneficial in general and essential for this specific domain of application.
A good uncertainty quantification framework should: (1) yield uncertainty estimates that are well correlated with the actual prediction errors; (2) avoid introducing a major computational overhead; (3) be easy to implement on top of arbitrary regression schemes; (4) allow for direct, accurate and robust uncertainty propagation to derived quantities.
In this work we discuss how different existing approaches fare with respect to these four criteria, and demonstrate how  a ``calibrated shallow ensemble" (a committee of models that share part of their weights, and estimate errors as the standard deviation of their predictions) provide a good tradeoff, being especially convenient to propagate errors when combining multiple model predictions. 

This work is structured as follows.
In Section~\ref{sec:uq-theory} we discuss the design space of uncertainty estimators, post-hoc calibration of uncertainty estimates and the propagation of uncertainty to composite quantities and averages. %
In Section~\ref{sec:uq-general} we introduce the direct propagation of shallow ensembles (DPOSE) approach, and compare systematically many of the possible combination of uncertainty estimation methods, using a well-established collection of 10 datasets to assess the impact of different choices in the context of general regression tasks.
Then, in Section~\ref{sec:uq-atomistic} we demonstrate  the viability of our approach for the prediction of molecular formation energies, and potential energy and forces of condensed-phase systems, namely liquid water, barium titanate, and the solid state electrolyte \ce{Li3PS4}, as well as the molecular dataset QM9. 
We highlight in particular how the structure of MLIPs requires models that are capable of reliable uncertainty propagation even to just consistently define the error in the total energy of a simulation. We demonstrate the propagation of uncertainties to the average structural and thermophysical properties of liquid water, and perform ad hoc computational experiments to assess the ability of a calibrated shallow ensemble model to quantify errors for out-of-distribution data.

\section{Uncertainty quantification}
\label{sec:uq-theory}

Uncertainty quantification is a very broad term describing algorithms and procedures aimed at assessing the error in a measurement or the simulation of a quantity of interest~\cite{abdarReviewUncertaintyQuantification2021b}.  
In the context of machine-learning regression models, UQ refers to procedures that attempt to determine the reliability of the values predicted by the model, or that of quantities derived from them. 
Potential sources of uncertainty include reducible \emph{epistemic} uncertainties due e.g. to finite sampling of the training data, or to extrapolation to unchartered portions of the input parameter space, and irreducible \emph{aleatoric} uncertainties such as noise in the reference targets~\cite{hullermeierAleatoricEpistemicUncertainty2021b}, or the errors associated with lack of descriptive power in the model. 
In the context of atomistic simulations, aleatoric uncertainty might be due to the use of a reference quantum calculation that has an intrinsic stochastic error (e.g. quantum Monte Carlo) while approximation errors include neglecting long-range interactions by focusing on short-range structural descriptors~\cite{nibl+21jcp,zhai+23jcp,huguenin-dumittanPhysicsinspiredEquivariantDescriptors2023}, or using an architecture and/or geometric descriptors lacking universal approximating capabilities~\cite{pozdnyakovIncompletenessAtomicStructure2020b,pozdnyakovIncompletenessGraphNeural2022}.

Many recent research papers are dedicated to uncertainty quantification in atomic-scale modeling~\cite{buskCalibratedUncertaintyMolecular2022,buskGraphNeuralNetwork2023a,thalerScalableBayesianUncertainty2023,tanSinglemodelUncertaintyQuantification2023a,pauljanetQuantitativeUncertaintyMetric2019,zhuFastUncertaintyEstimates2023,huRobustScalableUncertainty2022,musi+19jctc,carreteDeepEnsemblesVs2023,itzavazquez-salazarUncertaintyQuantificationPredictions2022,longbottomUncertaintyQuantificationClassical2019,rensmeyerHighAccuracyUncertaintyAware2023,venturiBayesianMachineLearning2020,wenUncertaintyQuantificationMolecular2020,zaverkinExplorationTransferableUniformly2021a,kahleQualityUncertaintyEstimates2022,wenUncertaintyQuantificationMolecular2020,a.petersonAddressingUncertaintyAtomistic2017,pernotPredictionUncertaintyValidation2022a,pernotCalibrationMachineLearning2023,bartokImprovedUncertaintyQuantification2022,annevelinkStatisticalMethodsResolving2023a,xieUncertaintyawareMolecularDynamics2023,yangExplainableUncertaintyQuantifications2023,thomas-mitchellCalibrationUncertaintyActive2023, tynesPairwiseDifferenceRegression2021, scaliaEvaluatingScalableUncertainty2020, breuckRobustModelBenchmarking2021}, which underscores the importance of the topic for the progress of this community. 
The design space of uncertainty models is quite broad (see Fig.~\ref{fig:uq-summary}) and includes: the model architecture (how the uncertainty is computed); the training target and protocol, and the closely-connected choice of the figure of merit used to assess the quality of the estimated errors; the calibration of the prediction; the propagation of uncertainty for derived quantities. 
In this Section we provide a concise but thorough overview of these different ingredients.

\subsection{Uncertainty model architecture}
Gaussian process regression (GPR) is a paradigmatic example of a probabilistic modelling scheme that directly gives access to uncertainty estimates. In GPR the mean and uncertainty of the target property $y$ of a sample $A$ are given by the closed form expressions:
\begin{equation}
    \label{eq:Kernel_UQ}
 \bar{y}(A) = \boldsymbol{\alpha}\vb{k}(A) \quad    \sigma^2(A) = k - \vb{k}^T \vb{K}^{-1}\vb{k}.
\end{equation}
The vector $\vb{k}(A)$ contains the covariances or kernel similarities between the new test sample $A$ and the training points, $k$ is the covariance of the new training point with itself. $\vb{K}^{-1}$ is the inverse kernel matrix of the training points.
Evaluating uncertainty through Eq.~\eqref{eq:Kernel_UQ} is computationally demanding, even when the Nystroem approximation is used and the inverted matrix is precomputed~\cite{rasmussenGaussianProcessesMachine2008}.

Besides the computational cost, Eq.~\eqref{eq:Kernel_UQ} is limited to the case of GPR models. 
An approach that can be used to incorporate an error estimate for non-linear models is mean variance estimation (MVE), that uses a neural network to parameterize both the mean $\bar{y}_{\theta_{\textrm{NN}}}(A)$ and the variance $\sigma_{\theta_{\textrm{NN}}}^2(A)$,  interpreted as the parameters of a normal distribution of the model predictions
\begin{equation}
    p_{\theta_{\textrm{NN}}}(y|A) = \mathcal{N}(y|\bar{y}_{\theta_{\textrm{NN}}}(A),\sigma_{\theta_{\textrm{NN}}}^2(A))
\end{equation}
Mixture density networks, or deep evidential regression are closely related, and employ neural networks to parameterize more flexible probability distributions~\cite{williams1996using,aminiDeepEvidentialRegression2020a}.

A generally-applicable and well-established strategy for uncertainty quantification involves defining an 
\emph{ensemble} (or commitee) of $\nens$ models~\cite{epsteinStochasticDynamicPrediction1969, epsteinRoleInitialUncertainties1969, rafteryUsingBayesianModel2005, tothEnsembleForecastingNMC1993,gneitingCalibratedProbabilisticForecasting2005, lakshminarayananSimpleScalablePredictive2017a}.
In ensembles, the mean of the committee predictions $\bar{y}(A)=\sum^{\nens}_{k=1}y^{(k)}(A)/\nens$ is taken as the best estimate of the model, while the uncertainty is expressed as the variance $\sigma_{\nens}^2(A)$ of the individual committee member predictions
\begin{equation}
    \sigma^2_{\nens}(A) = \frac{1}{\nens - 1} \sum^{\nens}_{k=1}{[y^{(k)}(A) - \bar{y}(A)]^2}.
\end{equation}
\rev{
While it is common to assume the error distribution to be Gaussian (e.g. invoking the central limit theorem) the ensemble formalism allows defining a non-parametric estimate of the predicted distribution 
\begin{equation}
p(y|A)=\sum_k \delta(y-y^{(k)}(A)), \label{eq:ensemble-nonpara}
\end{equation}
where the $\delta$ distribution is usually smoothed into finite-width Gaussian peaks.
}

An architecture that combines ideas from MVE and ensemble models is the ``deep ensemble'' framework introduced by  Lakshminarayanan and coworkers~\cite{lakshminarayananSimpleScalablePredictive2017a}. 
Each member of the committee is effectively a MVE model that evaluates a prediction $y^{(k)}(A)$  and a variance ${\sigma^2}^{(k)}(A)$ - which is why in Fig.~\ref{fig:uq-summary} we label this architecture with the more expressive name ``mean-variance ensemble''.
The uncertainty estimate is obtained as the sum of the mean of the individual model prediction variances and the variance of the means of the individual variance predictions.

\begin{equation}
\sigma_{\textrm{tot}}^2(A) = \frac{1}{\nens - 1} \sum^{\nens}_{k=1}{[y^{(k)}(A) - \bar{y}(A)]^2} + \frac{1}{\nens} \sum^{\nens}_{k=1} {\sigma^2}^{(k)}(A)
\end{equation}
Busk et al. used deep ensembles to obtain reliable uncertainty estimates for the property prediction of small organic molecules~\cite{buskCalibratedUncertaintyMolecular2022, buskGraphNeuralNetwork2023a}.
Thaler and coworkers observed that deep ensembles yield uncertainty quantifications of comparable quality to using stochastic gradient Markov chain Monte Carlo to sample the posterior~\cite{thalerScalableBayesianUncertainty2023}.

Ensembling of fully-independent neural-network models introduces a significant computational overhead, as it requires training and evaluating multiple models, whereas  approaches such as mean variance estimating neural networks and deep evidential regression that directly predict both a mean and its variance, with a single neural network, are usually less demanding~\cite{nixEstimatingMeanVariance1994,aminiDeepEvidentialRegression2020a, tanSinglemodelUncertaintyQuantification2023a}.
Multiple recent efforts have attempted to reduce the computational cost of uncertainty estimation, 
by relying on distance based criteria \cite{pauljanetQuantitativeUncertaintyMetric2019}, Gaussian Mixture models on latent features \cite{zhuFastUncertaintyEstimates2023} or conformal prediction \cite{huRobustScalableUncertainty2022}.
Another approach is to reduce the cost of ensemble methods by sharing parts of the computational graph between committee members~\cite{leeWhyHeadsAre2015}. 
For instance, in Ref.~\citenum{carreteDeepEnsemblesVs2023}, fully independent ensembles of models  are trained, but most of the computational time in their architecture is spent on calculating features, which are shared by all models. 
Finally, sometimes the members of an ensemble model are chosen to have slight variations in their architecture, e.g. by modifying some hyperparameters, or by changing the number of layers and/or neurons in an ensemble of multi-layer perceptrons~\cite{zhuFastUncertaintyEstimates2023}. In this work, we will restrict ourselves to ensembles that share the same architecture and hyperparameters.

\begin{figure*}[bthp]
\includegraphics[width=1.0\linewidth]{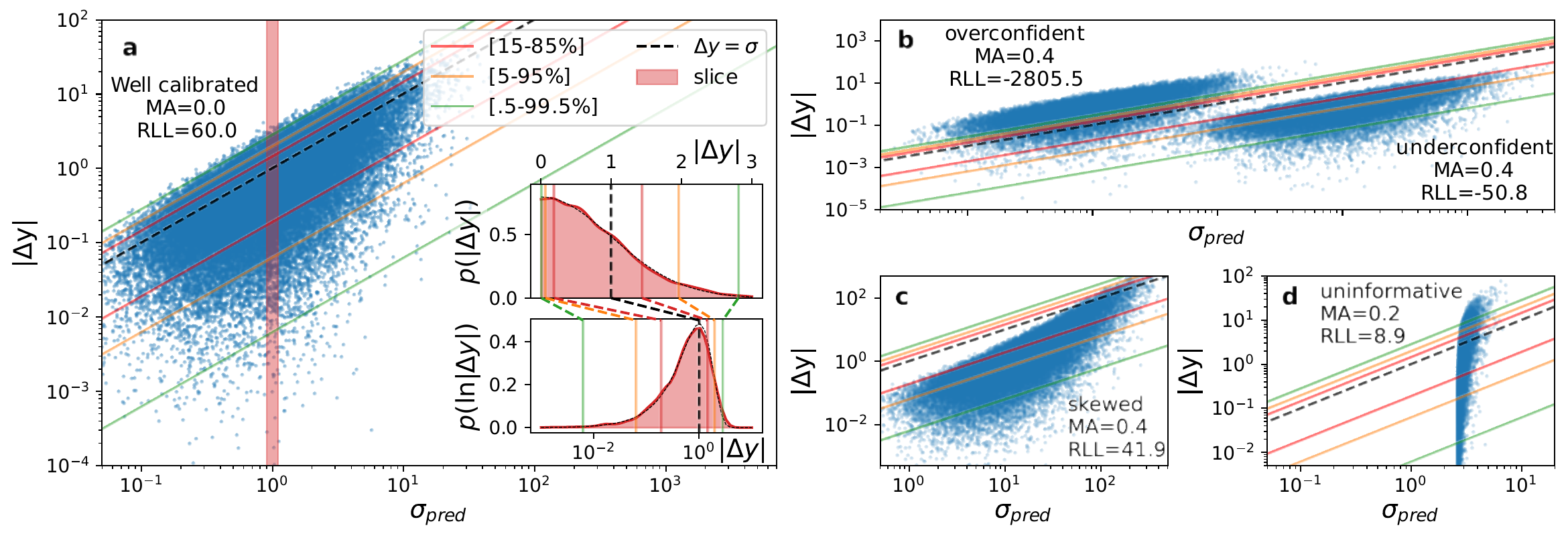}
\caption{\rev{
Schematic overview of different modes of failure of uncertainty quantification models, based on a toy model whose predictions for an input $x$ are affected by a purely aleatoric Gaussian error with a standard deviation proportional to $x$. 
(a) Predicted-empirical error scatter plot: we will use this type of visualization as the main tool to characteriza uncertainty models; due to the log-log scale, the Gaussian distribution of empirical error deforms into an asymmetric bell shape, with the dashed black line corresponding to the mode for a well-calibrated prediction, and the colored lines corresponding to different quantile ranges, that are useful as guides for the eye. The insets show the distribution for a slice of the dataset with predicted uncertainty close to $1$, and serve to emphasize the change in appearance of the distribution when using a log scale for the empirical error.  
(b) Examples of poorly calibrated estimators, together with their miscalibration area (MA) and standardized log-likelihood (RLL). Note that the RLL is not symmetric with respect to miscalibration, and is more sensitive to overconfident than underconfident predictions. 
(c) Example of an estimator that is globally calibrated (the mean predicted variance matches the MSE over the test set), but which is locally mis-calibrated.
(d) Example of an estimator that is well calibrated but poorly informative, as all structures are predicted to have roughly the same error. 
}}
\label{fig:toy}
\end{figure*}

\subsection{Assessing the quality of UQ}

The accuracy of a ML model is typically checked by computing the error of its predictions, e.g. through the mean squared error (MSE) $|y(A) - y_\text{ref}(A)|^2$  or the mean absolute error (MAE) $|y(A) - y_\text{ref}(A)|$, summed over the validation  (or test) set. 
When assessing a model that computes both a best estimate $\bar{y}(A)$ and an uncertainty $\sigma(A)$, a proper metric should consider the accuracy in predicting both, penalizing under or over-confident uncertainty estimates. 
If a model provides a probability distribution for its predictions $p(y|A)$, then one way to assess the quality of the predicted distribution is to compute the probability (likelihood) $p(y_\text{ref}(A)|A)$ for a hold-out set of validation samples. 
In practice one usually expresses this information in terms of the negative log-likelihood $\text{NLL}=-\ln p(y_\text{ref}(A)|A)$, where the logarithm means that the joint probability for all samples in a validation set can be computed by \emph{summing} the NLL values, and the minus sign ensures that ``good'' values are small, as for the MSE or the MAE. 
Under the assumption of a Gaussian probability (that is implicit in measuring uncertainty using only the second moment of $p$) the NLL can be written as 
\begin{equation}
\label{eq:nll}
\text{NLL}(\Delta{y},\sigma) = \frac{1}{2} \left[\frac{\Delta y^2}{\sigma^2} + \ln 2\pi \sigma^2\right],
\end{equation}
that depends on the prediction error $\Delta y = y-y_\text{ref}$ and the uncertainty $\sigma$.
While it is possible to use the NLL to compare two models over the same data set, its absolute magnitude is not as transparent as the MAE or MSE for the prediction error.

Rasmussen and Williams proposed to standardize the NLL (in order to compare predictions across different datasets) by substracting the likelihoods of a model predicting mean and variance of the training dataset~\cite{rasmussenGaussianProcessesMachine2008}. 
We further normalize the scale of variation of the NLL defining  a \emph{relative} log-likelihood (RLL) that is computed  as 
\begin{equation}
\frac{
\sum_{A}  \text{NLL}(\Delta y(A), \sigma(A)) - \text{NLL}(\Delta y(A), \text{RMSE})
}{
\sum_{A} 
\text{NLL}(\Delta y(A), |\Delta y(A)|) - 
\text{NLL}(\Delta y(A), \text{RMSE}) 
}\cdot 100\%.
\label{eq:snll}
\end{equation}
As in Ref.~\citenum{rasmussenGaussianProcessesMachine2008}, the numerator measures the difference between the actual NLL and a ``calibrated constant error'' estimate in which every structure is predicted to have the same error, equal to the root mean square validation error -- which is the least informative  uncertainty estimate. 
The denominator involves an ``oracle'' NLL that predicts for each sample an uncertainty that is equal to the unsigned error (the lowest NLL for a given prediction error). 
With this definition, larger values of the RLL are better, and a value of 100\%{} would indicate the best possible score, which is however in practice almost impossible to achieve, as predicting the unsigned error should be almost as difficult as predicting the actual error and correcting the model estimate. 
A value of zero indicates that -- estimates $\bar{y}(A)$ being the same -- the UQ is uninformative, equivalent to using a constant value that equals the overall RMSE. Negative values indicate poor calibration of the model.
The RLL is easier to interpret than the NLL, and less dependent on the model accuracy, as it is ``baselined'' on the NLL computed for the model RMSE. 
Even though the NLL (and the equivalent, but more easily interpretable, RLL) provide a holistic assessment of the reliability of an error estimate, it is often useful to understand \emph{how} an estimator may be failing -- which has led to the development of a multitude of diagnostic~\cite{gneitingStrictlyProperScoring2007,chungUncertaintyToolboxOpenSource2021a,pern22jcp}.
To illustrate some of these metrics we use a toy example, corresponding to a random variable $y(x)$, sampled from a normal distribution with zero mean and standard deviation proportional to $x$.
\rev{
To obtain a realistic distribution of points, the inputs $x$ are taken to be log-normal distributed, with mean value of $1$. 
We visualize the behavior of the uncertainty estimator on a validation set using a scatter plot that shows the unsigned error of a prediction (the residual $|\Delta y(x)| = |\bar{y}(x)-y_\text{ref}(x)|$) against the predicted standard deviation $\sigma(x)$ (Fig.~\ref{fig:toy}a).
When interpreting these plots, it is important to keep in mind that, for a typical Gaussian profile of the prediction distribution $p(y|A)$, the most likely error is \emph{zero} regardless of the uncertainty. 
Furthermore, in order to resolve the behavior of the uncertainty estimator over a broad range of values, it is useful to use a doubly-logarithmic scale for the plot. 
This distorts the error distribution, that becomes asymmetric even in the Gaussian case, with a maximum at $\Delta y=\sigma$. 
Panel (a) in Fig.~\ref{fig:toy} provides a key to interpret this type of scatter plots. The distribution of points along the $x$ axis reflects the heteroskedactic nature of the error: in this case, most points accumulate around $\sigma=1$, with log-normal tails for smaller and larger errors. 
If one selects a thin slice of values for the predicted uncertainty, the absolute errors for the selected points are half-normal-distributed, consistent with the dataset construction. 
When plotted with a logarithmic $|\Delta y|$ axis, the half normal distribution is morphed in an asymmetric peak shape, with a maximum at $\Delta y=\sigma$, a sharp decay for $\Delta y > \sigma$ a long tail for small error values. As a guide for the eye, we plot -- here and in all similar plots in this paper -- the quantile lines expected for this distribution.
Thus, a good uncertainty estimator should yield, for each vertical slice along the $\sigma$ axis, a distribution that is reminiscent to that shown in the insets of panel (a). 

Conversely, this scatter plot allows identifying the different modes of failure of an uncertainty estimator.
Systematic underestimation or overestimation of the residuals indicate poor \emph{calibration} of the estimator, which is said to be over(under)confident, and is visible in a rigid shift of the distribution that is not peaked at $\Delta y =\sigma$ (e.g Fig.~\ref{fig:toy}b). }

In regression problems, this type of failure can be identified by several metrics, such as the expected normalized calibration error~\cite{leviEvaluatingCalibratingUncertainty2020a}, the miscalibration area~\cite{tranMethodsComparingUncertainty2020} or the $z$-score variance~\cite{pernotPredictionUncertaintyValidation2022a}. 
We will use the \emph{miscalibration area} (MA), that measures how much the empirical cumulative distribution function (CDF) of the predictions over a dataset deviates from that derived from a probabilistic uncertainty model $p(y|x)$. 
In practice, we use the standardized residuals $\Delta y(x)/\sigma(x)$ when constructing the empirical  CDF, which is then compared to the CDF of the unit Gaussian -- so that a MA of zero indicates that \emph{on average} residuals are as large as predicted, whereas values larger than zero indicate that the uncertainty estimates are under or over confident. 
One simple way of defining the overall calibration of an estimator is to compare the mean square error predicted over a validation set with the empirical mean square error: as discussed in subsection~\ref{sub:post-hoc}, it is then easy to obtain a calibrated model for which the two are equal, by a simple rescaling of the predicted error. 
The overall calibration, however, gives only partial  information on the behavior of an estimator, which can be calibrated as a whole, but over(under) confident for some subsets of the dataset. \rev{This scenario is clearly identified by having subsets of the data points that are consistent with the quantile lines and others that have an incorrect offset (see e.g. Fig.~\ref{fig:toy}c).} 
The NLL, as well as many commonly-used diagnostics, are sensitive to this kind of errors. Alternatively, to gather more granular insights, one can also compute calibration metrics for subsets of the overall data -- e.g. stratifying the evaluation of the z-score according to the predicted $\sigma(x)$, as done e.g. in evaluating the local-z-score variance ~\cite{, pernotPredictionUncertaintyValidation2022a,pernotCalibrationMachineLearning2023}.
\rev{Finally, an uncertainty estimator can also be well-calibrated but \emph{uninformative}, because it cannot distinguish predictions with low and high uncertainty (cf. Fig.~\ref{fig:toy}d).  }
An extreme case is one in which the same uncertainty is predicted for every data point (the constant-error baseline used in the definition of the RLL), but more generally it is preferable to have an estimator that yields a broad range of uncertainty values, distinguishing between reliable and unreliable predictions.
One can confirm that an estimator predicts a broad range of uncertainties (the \emph{dispersion} of the estimator) by evaluating its coefficient of variation~\cite{leviEvaluatingCalibratingUncertainty2020a}.
Many other metrics have been proposed to assess the quality of error estimators, such as sharpness~\cite{gneitingProbabilisticForecastsCalibration2007},  the prediction interval coverage probability~\cite{guoCalibrationModernNeural2017, kuleshovAccurateUncertaintiesDeepb} and many others \cite{gneitingStrictlyProperScoring2007,pernotCalibrationMachineLearning2023}. 
Overall, even though they can provide complementary diagnostics for the relative merits of different approaches, we find most of these metrics not to be particularly insightful or easy to interpret, and we will mainly rely on uncertainty-residual parity plots as a way to visualize the behavior of an estimator, and on the RLL as a single metric that assesses quantitatively its overall performance. 

\subsection{Training strategy}

A regression model is usually optimized by minimizing the error it incurs when predicting the target properties of a set of samples for which the ground truth property values are known (the training set). 
When optimizing an architecture that includes an error estimator, there are different strategies one can follow (cf. Fig.~\ref{fig:uq-summary}). 
First, some regression models (e.g. Gaussian processes) can be trained based exclusively on the mean square error of the mean prediction $\bar{y}$, and yet provide an estimate for $\sigma$. This often leads to poor calibration of the predicted variance, and require further tuning, as discussed in the next subsection. 
For ensemble models, it is possible to train separately the various committee members: by varying the initialization of the weights, the hyperparameters, or by randomly subsampling the training set\cite{musi+19jctc}, the models will converge to a different local minimum, and their spread can be used (possibly following further calibration) to estimate $\sigma$.
Finally, one can incorporate the metrics discussed in the previous paragraph into the training loss. 
By optimizing the NLL, for instance,  one can make sure that not only the predicted mean $\bar{y}$ but also the standard deviation $\sigma$ are consistent with the empirical distribution of the training data -- improving calibration. 
This approach can be applied to any UQ architecture, including models that provide an explicit estimate for $\sigma$, even though it has been sometimes found to cause instabilities in the learning procedure~\cite{seitzerPitfallsHeteroscedasticUncertainty2021}.

\subsection{Post-hoc calibration}
\label{sub:post-hoc}

In all cases in which an uncertainty-quantified models is mis-calibrated (e.g. for ensembles of neural networks that tend to produce overconfident uncertainty estimates\cite{kuleshovAccurateUncertaintiesDeepb,
guoCalibrationModernNeural2017,
clarteDoubledescentUncertaintyQuantification2023, clarteTheoreticalCharacterizationUncertainty2023}), %
it is possible to apply a post-hoc calibration step on a hold-out set (in practice we use the validation set) to globally correct a model's uncertainty estimates.
In the simplest case, assuming that the target properties are Gaussian distributed, one can apply a simple rescaling $\sigma\leftarrow \alpha \sigma$. 
The constant $\alpha$ that minimizes the Gaussian NLL on the calibration can be derived from the squared residuals $\Delta y(A)^{2}$ and the predicted variances $\sigma(A)^2$:
\begin{equation}
    \alpha^{2} = \frac{1}{n_{\textrm{val}}}\sum_{A\in\text{val}}\frac{\Delta y(A)^{2}}{\sigma(A)^{2}}.
\end{equation}
When using ensemble methods with small $\nens$, a correction may be used to account for the correlation between $\bar{y}(A)$ and $\sigma(A)$, as discussed in Ref.~\citenum{imba+21jcp}.

More sophisticated calibration strategies allow to correct for local mis-calibration, e.g. using a non-linear transformation of the predicted variance, e.g. optimizing an isotonic regressor~\cite{kuleshovAccurateUncertaintiesDeepb, buskCalibratedUncertaintyMolecular2022}. %
Implementations of these calibration frameworks can be found in the netcal library \cite{Kueppers_2022_ECCV_Workshops} and the uncertainty toolbox library\cite{chungUncertaintyToolboxOpenSource2021a}.
An important observation that we will use in what follows is that, in the case of ensemble estimators, one can apply the calibration mechanism to the members of the committee rather than to $\sigma$. If $\alpha(A)$ is the (possibly sample-dependent) calibration factor, such \emph{calibrated ensemble}\cite{musi+19jctc}, \rev{consistent with the empirical error distribution,}
  can be obtained as
\begin{equation}
y^{(k)}(A)\leftarrow \bar{y}(A) + \alpha(A) [ y^{(k)}(A)- \bar{y}(A)].
\label{eq:calibrated-ensemble}
\end{equation}

\subsection{Uncertainty propagation}

In many cases the predictions of a ML model are not the ultimate target of a calculation, but are subject to further manipulations, either by combining multiple predictions or by applying some function $f$ to the output of the model, 
\begin{equation}
    z(A) = f(y(A)).
\end{equation}
In these cases, it becomes necessary to determine how the uncertainties of the model predictions $\sigma_{y}(A)$ propagate to the uncertainty of $z$, $\sigma_{z}(A)$.
The most straightforward approach is to apply Gaussian error calculus, linearizing $f$ to estimate the effect of the uncertainties of the variables of $f$ on $\sigma_{z}$, i.e.  $\sigma_z(A) \approx \sigma_y(A) f'(y(A))$.
Additional complications may arise if a linear approximation of $f$ is inaccurate, input variables are correlated or, for practical reasons, analytical derivatives of $f$ are not available. Monte Carlo approaches to estimate the uncertainty of $z$ can be a suitable alternative, implying however that $f$ has to be evaluated repeatedly~\cite{zhangModernMonteCarlo2021}.
Within the ensemble formalism, propagating the uncertainty of $y(A)$ from individual model predictions $y^{(k)}(A)$ to $z(A)$ can be performed by treating the individual committee member predictions as samples drawn from the distribution $p(y|A)$ and then evaluating $\sigma^2_{z}$ from samples of $f(y^{(k)}(A))$ (cf. Fig.~\ref{fig:toy})
\begin{equation}
\begin{split}
\bar{z}(A) =& \frac{1}{\nens} \sum^{\nens}_{k=1}f(y^{(k)}(A))), \\
\sigma^2_z(A) =& \frac{1}{\nens - 1} \sum^{\nens}_{k=1}{[f(y^{(k)}(A)) - \bar{z}(A)]^2}.
\end{split}\label{eq:ensemble-propagation}
\end{equation}
One sees the importance of using ensemble calibration~\eqref{eq:calibrated-ensemble} to ensure that the committee members $y^{(k)}(A)$ reflect as well as possible the correct error distribution $p(y|A)$. We refer to Eq.~\eqref{eq:ensemble-propagation} as \emph{ensemble propagation}.

One particular case, that occurs frequently in atomistic simulations but is also relevant for generic applications of Metropolis-Hastings sampling \cite{hitchcockHistoryMetropolisHastings2003, hamraMarkovChainMonte2013}, is that in which a sequence of samples $\{A_i\}_{i=1}^N$ is generated using a machine-learning potential energy function $V(A)$, and the average value for a property $y(A)$ is computed by summing over these samples, $\left<y\right>_V = N^{-1} \sum_i y(A_i) $. 
The samples are generated with probabilities $\propto e^{- \beta V(A_i)}$ -- so that errors in predicting $V(A)$ will propagate onto the average $\left<y\right>_V$ by altering the sample distribution. 
If $V$ is computed with an ensemble method, one could perform separate sampling trajectories using the different members of the ensemble, which however would substantially increase the computational effort in determining $\left<y\right>$. 
Alternatively, as discussed in Ref.~\citenum{imba+21jcp}, one can use \emph{statistical reweighting} to obtain an estimate of the impact of the error on $V$ on the distribution of the samples, using a single set of samples distributed according to the mean ensemble potential $\bar{V}(A)$
\begin{equation}
\label{eq:reweight-cea}
\left<y\right>_{V^{(k)}} \approx \expval{y}_{\bar{V}} - \beta [\expval{y(V^{(k)}-\bar{V})}_{\bar{V}} - \expval{y}_{\bar{V}}\expval{V^{(k)}-\bar{V}}_{\bar{V}}].
\end{equation}
This expression arises from averaging over a re-weighted trajectory~\cite{torr-vall99jcp}, which is approximated by a cumulant expansion to avoid statistical instability when the deviations between the committee members and the mean potential are large~\cite{ceri+12prsa}.

Error propagation through ensembles is also useful to estimate errors on derivatives of the model target with respect to a continuous parameter $\lambda$ that characterizes the samples -- e.g. forces acting on atoms, that are derivatives of the potential with respect to atomic coordinates. 
Defining derivative errors based on a direct variance estimator is not trivial\cite{carreteDeepEnsemblesVs2023}, whereas a calibrated ensemble allows to do so straightforwardly, as we shall discuss later.
\begin{equation}
\begin{split}
\bar{y}'(A) = & \frac{1}{\nens} \sum_{k=1}^{\nens} \frac{\partial{y^{(k)}(A)} }{\partial{\lambda}}  \\
\sigma_{y'}(A) = &   \frac{1}{\nens - 1} \sum^{\nens}_{k=1}{\left[\frac{\partial{y^{(k)}(A)} }{\partial{\lambda}} - \bar{y}'(A) \right]^2}.
\end{split}
\end{equation}

\renewcommand{\topfraction}{0.9}  %
\renewcommand{\bottomfraction}{0.8}  %
\renewcommand{\textfraction}{0.1}  %
\renewcommand{\floatpagefraction}{0.8}  %

\begin{figure}[tbp]
    \centering
    \includegraphics[width=0.7\linewidth]{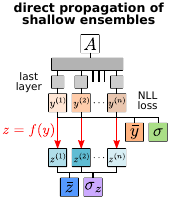}
    \caption{A schematic overview of the DPOSE architecture. The last layer of the underlying model architecture is split in a shallow ensemble of models, each making a separate prediction. 
    The mean and standard deviation of the ensemble values are used to define a NLL loss, which is used to train a calibrated ensemble. 
    Propagation of uncertainty on any derived quantity is obtained by separately applying the desired transformation to yield an ensemble of transformed quantities. }
    \label{fig:dpose}
\end{figure}

\section{Direct propagation of shallow ensembles}
\label{sec:uq-general}

\begin{table*}
\begin{tabular}{l @{\hskip 1mm} ccc @{\hskip 5mm} cccccc}
\toprule
Dataset       & MVE & MVEns & \textbf{DPOSE} & A    & B    & C    & D    & E    & F    \\ \midrule
Training      & NLL & NLL   & NLL           & NLL  & NLL  & NLL  & MSE  & MSE  & MSE-SS \\
$\nens$       & -   & 5     & 64            & 5    & 5    & 5    & 5    & 5    & 5     \\
Weight sharing& -   & no    & yes           & yes  & no   & yes  & no   & no   & no   \\
Post-hoc cal. & yes  & yes    & yes           & yes  & yes  & no   & no   & yes  & yes   \\ \midrule
Housing & $0.7^{\pm 16.7}$ & $7.1^{\pm 17.6}$ & $13.8^{\pm 20.1}$ & $-14.1^{\pm 41.8}$ & $-7.7^{\pm 35.4}$ & $-6.2^{\pm 34.0}$ & $-388^{\pm 296}$ & $-26.2^{\pm 43.3}$ & $-10.8^{\pm 25.9}$ \\
Concrete & $9.2^{\pm 16.3}$ & $17.1^{\pm 8.1}$ & $18.6^{\pm 12.4}$ & $-2.3^{\pm 18.4}$ & $-10.5^{\pm 32.9}$ & $-1.9^{\pm 26.4}$ & $-697^{\pm 362}$ & $-19.2^{\pm 23.3}$ & $-12.2^{\pm 13.6}$ \\
Energy & $44.7^{\pm 6.3}$ & $48.0^{\pm 4.5}$ & $51.3^{\pm 7.8}$ & $26.6^{\pm 9.5}$ & $21.1^{\pm 12.5}$ & $13.8^{\pm 16.5}$ & $-449^{\pm 203}$ & $-14.4^{\pm 15.6}$ & $-5.9^{\pm 22.9}$ \\
Kin8nm & $14.9^{\pm 3.6}$ & $17.7^{\pm 2.5}$ & $13.5^{\pm 4.1}$ & $13.4^{\pm 3.5}$ & $12.7^{\pm 3.6}$ & $-2.8^{\pm 21.7}$ & $-283^{\pm 83}$ & $-27.3^{\pm 9.9}$ & $-23.5^{\pm 7.9}$ \\
Naval & $46.1^{\pm 11.7}$ & $44.9^{\pm 9.9}$ & $43.8^{\pm 9.9}$ & $29.2^{\pm 19.3}$ & $6.6^{\pm 36.9}$ & $25.9^{\pm 51.9}$ & $12^{\pm 16}$ & $19.6^{\pm 10.2}$ & $17.0^{\pm 8.5}$ \\
Power & $1.1^{\pm 2.1}$ & $1.9^{\pm 2.4}$ & $0.8^{\pm 3.2}$ & $-3.1^{\pm 15.5}$ & $-4.2^{\pm 17.0}$ & $-7.6^{\pm 9.5}$ & $-3571^{\pm 1569}$ & $-41.6^{\pm 17.6}$ & $-37.9^{\pm 11.2}$ \\
Protein & $17.1^{\pm 9.6}$ & $13.7^{\pm 2.0}$ & $13.2^{\pm 5.2}$ & $6.5^{\pm 8.4}$ & $6.1^{\pm 8.6}$ & $9.4^{\pm 32.5}$ & $-4120^{\pm 1576}$ & $-51.6^{\pm 20.8}$ & $-55.1^{\pm 10.5}$ \\
Wine & $-5.2^{\pm 10.8}$ & $2.2^{\pm 7.2}$ & $-5.7^{\pm 18.5}$ & $-8.5^{\pm 11.9}$ & $-16.6^{\pm 21.1}$ & $-15.4^{\pm 22.4}$ & $-1776^{\pm 562}$ & $-32.7^{\pm 17.2}$ & $-36.0^{\pm 16.6}$ \\
Yacht & $3.4^{\pm 32.1}$ & $33.3^{\pm 11.8}$ & $64.5^{\pm 18.6}$ & $44.7^{\pm 22.8}$ & $37.6^{\pm 29.1}$ & $53.1^{\pm 15.3}$ & $-382^{\pm 511}$ & $-21.3^{\pm 57.9}$ & $4.8^{\pm 19.4}$ \\
Years & $20.5 $ & $25.1$ & $23.6$ & $23.6$ & $22.9$ & $20.8$ & $-2825$ & $-16.8$ & $-16.8$ \\
\bottomrule
\end{tabular}

\caption{Performance of different UQ approaches for a set of regression benchmarks\cite{lakshminarayananSimpleScalablePredictive2017a,hernandez-lobatoProbabilisticBackpropagationScalable2015}. 
To facilitate comparison between datasets, the UQ performance is quantified by the relative log likelihood (RLL, eq.~\eqref{eq:snll}, higher is better); a comparison with literature results using the plain NLL are shown in the SI. 
Standard deviations over multiple train/test splits are shown for each value, except for the larger ``years'' dataset, for which we use a single split. 
MVE and MVEns indicate a mean-variance estimator architecture, and a mean-variance deep ensemble; DPOSE indicate the proposed direct propagation of shallow-ensembles approach; models A-F are several variations on a theme for DPOSE, differing by the training target (NLL, or MSE for separate models), by the number of committee members $\nens$, by the use of weight sharing for all but the last network layer, by the use of post-hoc calibration on a hold-out set. 
\label{tab:general-benchmarks-RLL}}
\end{table*}

In Section~\ref{sec:uq-theory} we have provided an opinionated overview of the vast space of design choices available for uncertainty-quantified ML models.
In this section we intend to demonstrate the behavior of a specific set of choices, that we will refer to as ``direct propagation of shallow ensembles'' (DPOSE), and argue that it fulfils the requirements of accuracy, ease of implementation, low computational overhead and, in particular, simplicity of error propagation. 
DPOSE combines (1) an ensemble architecture that outputs a collection of target predictions, based on which mean and variance of the ensemble are computed; (2) weight sharing up to the last hidden layer, that reduces the computational overhead and simplifies implementation; (3) training by minimization of the NLL, reducing the need for post-hoc calibration; (4) error estimation on derived quantities through ensemble propagation (see also Figure~\ref{fig:dpose}).
In practice this formulation can be applied to most deep learning architectures and requires only modifiying the last layer weights that map the latent feature on the target properties and the loss function. The computational overhead is similar to adding an additional hidden layer to the neural network.

\subsection{Benchmarks on general datasets}

DPOSE models exhibit comparable performance to other established uncertainty-quantified architectures on a collection of benchmark datasets
~\footnote{The original collection of datasets, available from Refs.~\cite{kellyuci, DelveKinFamily, StatLibArchive}, also include results for the Boston housing
dataset, which has since been deprecated because of the
presence of racial biases. We include them for consistency with the original publication, as we use them for applications with no ethical or policy implications. }, the regression subset of the examples introduced by Lakshminarayanan et al. and Hernández-Lobato \cite{lakshminarayananSimpleScalablePredictive2017a,hernandez-lobatoProbabilisticBackpropagationScalable2015}.
We use the same model architecture and parameters in the original experiments, using a single-layer perceptron with 50 hidden neurons for the smaller datasets and 100 neurons for the larger datasets (``protein'' and ``years''), with ReLU activation functions. 
We repeat each experiment 20 times for the smaller datasets, 5 times for the larger protein dataset and once for the largest years dataset, employing the dataset-specific splits, when applicable. 
We reproduce the findings from Refs.~\citenum{lakshminarayananSimpleScalablePredictive2017a,hernandez-lobatoProbabilisticBackpropagationScalable2015} if we reduce the fixed learning rates of the experiments to $10^{-2}$ ($10^{-3}$ for the years dataset), a difference we attribute to different default weight initializations of the software packages.
To allow applying post-hoc calibration, we performed our tests with a 10\%{} hold-out set, leading to a small degradation of the results relative to those found in the literature (results with the 90-10 split used in previous works, showing RMSE and NLL as metrics, are reported in the SI).
Table~\ref{tab:general-benchmarks-RLL} compares the performance of a simple mean-variance estimator architecture, of a mean-variance ``deep ensemble'', and of a set of different ensemble models, using the RLL to provide a more intuitive comparison between the various data set.  
A first observation is that a DPOSE model demonstrates similar performance to the reference architectures. No adverse effects are observed by not including an explicit variance estimator, nor using only the last layer weights to differentiate between ensembles. All models have (within the variability observed with multiple splits) a positive RLL, but for most datasets the performance is rather poor, in many cases being statistically equivalent the the naive RMSE estimator (that we recall corresponds to an RLL value of zero).  
It is worth noting that performance degrades slightly when using a smaller number of committee members (model A), which is of little concern given that weight sharing makes the cost of using a large $\nens$ negligible, and that convergence can be achieved with a smaller number of ensemble members (see SI). 
Right-hand columns in the table also show the small impact of several architectural choices on the model performance.
Using fully independent ensemble members without weight sharing (model B), does not improve model accuracy, and avoiding post-hoc calibration (model C) does not have statistically significant effects provided that the ensemble is trained based on the combined NLL. 
Training separate models on a $\ell^2$ loss (MSE, model D) with random weight  initialization, on the other hand, dramatically degrades the RLL. Applying post-hoc calibration (model E) improves the performance, underscoring the importance of calibration for methods that do not explicitly optimize a NLL\cite{musi+19jctc,kuleshovAccurateUncertaintiesDeepb, gneitingCalibratedProbabilisticForecasting2005}, but remains less performant than DPOSE or MVE-based models. This is also true when combining train set subsampling with separate training of multiple models based on an $\ell^2$ loss (MSE-SS, model F).
Overall, this systematic investigation of the design space of UQ models suggests that it is not necessary to include explicit variance estimators, that good accuracy can be achieved with a high level of weight sharing between the ensemble members, and that calibration -- either post-hoc or through explicit optimization of a NLL loss -- is necessary to achieve results that are better than a naive constant-error UQ scheme.

\section{Atomistic  modeling}\label{sec:uq-atomistic}

We now turn to the main area of application we consider, namely the use of ML to build surrogate models of the atomic-scale properties of matter. 
We focus in particular on the construction of \emph{interatomic potentials}, predictors of the stability of atomic configurations that can be used to generate structures that are consistent with the relevant thermodynamic boundary conditions. 

\subsection{Machine-learning potentials}
\label{sec:bpnn} 

Most empirical and machine learning models for property prediction in molecules and materials decompose the estimation of global properties (energy, electronic charge density, dielectric responses \ldots) into a sum of atomic contributions. 
Such a decomposition enhances the transferability of models among systems of varying sizes or atomic compositions and ensures consistent size-extensive predictions (e.g. the energy of two non-interacting molecules is the sum of the energy of the individual molecules).

The atomic contributions are obtained by applying either parametric or nonparametric models to fixed or learnable representations of \emph{local environments} $A_i$  centered around each atom $i$ in a structure $A$, comprising all neighbors within a set cutoff radius. The presence of a cutoff distance ensures linear scaling of the computational complexity of evaluating the model with the number of particles in a system~\cite{behl-parr07prl}.
\begin{equation}
\label{eq:atom-centered-v}
V(A) = \sum_{A_i \in A} v(A_i) = \sum_{A_i \in A}NN(\{\mathbf{r}_{ji}\}_{j\in A_i})
\end{equation}
In this work we use a combination of two well-established approaches as a paradigmatic example of the application of uncertainty quantification to atomic-scale modeling, even though the DPOSE architecture would be generally applicable. 
We use descriptors corresponding to symmetrized two and three-body correlations of the neighbor density (radial and power spectrum descriptors), using the smooth overlap of atomic positions (SOAP)\cite{bart+13prb} implementation in the \emph{rascaline} package \cite{frauxLuthafRascaline2024a}, that are invariant to rigid rotations of an environment, and feed them to a multi-layer perceptron to build an estimator of $v(A_i)$. 
The forces acting on the atoms can be calculated as the derivatives of the potential energy with respect to the atomic positions. 
This type of neural network architecture, that can be traced back to Behler and Parrinello (BPNN)\cite{behl-parr07prl}, is used in several widespread frameworks including ANI\cite{smit+17cs} and DeepMD\cite{wang+18cpc}.

We modify this architecture to implement a DPOSE estimator, by making each MLP output $\nens$ energy values, differing only by the last-layer weights. 
The ensemble model then predicts the energy as the mean of the ensembles, and the uncertainty as their spread; crucially -- as we shall further discuss later -- the error in the total energy is computed \emph{after} combining atomic predictions
\begin{equation}
\label{eq:committee-total-v}
\begin{split}
\bar{V}(A) = & \sum_{A_i\in A} \frac{1}{\nens} 
\sum_k v^{(k)}(A_i) =\frac{1}{\nens} 
\sum_k V^{(k)}(A)  \\
\sigma_V^2(A) = & \frac{1}{\nens -1}\sum_k \left[ \sum_{A_i\in A} v^{(k)}(A_i) -\bar{V}(A) \right]^2.
\end{split}
\end{equation}
Forces and force errors can be readily computed based on the derivatives of the committee members
\begin{equation}
\label{eq:committee-total-f}
\begin{split}
\bar{f}_{i\alpha} (A) =  & \frac{\partial \bar{V}(A)}{\partial r_{i\alpha}} = \frac{1}{\nens} 
\sum_k \frac{\partial V^{(k)}(A)}{\partial r_{i\alpha}}  \\
\sigma_{f_{i\alpha}}^2(A) = & 
\frac{1}{\nens -1}\sum_k 
\left[\frac{\partial V^{(k)}(A)}{\partial r_{i\alpha}} -\bar{f}_{i\alpha}(A) \right]^2.
\end{split}
\end{equation}
Eqs.~\eqref{eq:committee-total-v} and \eqref{eq:committee-total-f} can be seen as special cases of the ensemble propagation formula~\eqref{eq:ensemble-propagation}, that makes it extremely simple to define errors on forces and total energies even though the model predicts only local energy contributions.

We optimize all ensemble members simultaneously, using the mean and variance of the ensemble predictions in an NLL-like loss, in which each train structure contributes a term
\begin{equation}
\begin{split}
\ell^2(A) = &(1 - \lambda)\frac{1}{2}\left[\log(\sigma^2_{V}(A)) + \frac{\Delta V(A)^2}{\sigma^2_{V}(A)} \right] \\ & + \frac{\lambda}{3N_A}\sum_{i=1}^{N_A} \sum_{\alpha \in \{x,y,z\} } (\Delta f_{i\alpha} )^2,
\end{split}
\label{eq:atom-centered-loss}
\end{equation}
where $\Delta V = \bar{V} - V_\text{ref}$ and $\Delta f_{i\alpha} = \bar{f}_{i\alpha} -f_{i\alpha, \text{ref}} $ are the residuals for total potential, and the forces for each of the $N_A$ atoms in the structure. 
The hyperparameter $\lambda$ determines the relative importance of the NLL-like potential energy loss, and of the MSE-like force loss. It would be possible to include a NLL-like term for the force, but we observe that it is not needed to achieve good calibration of the force errors. 
Let us stress that given that only the total potential and forces are well-defined, the decomposition in~\eqref{eq:atom-centered-v} is somewhat arbitrary\cite{chon+23jctc}. Both the energy predictions and the error estimates need only be meaningful when computed for an entire structure, making it very important to have a consistent approach to link the uncertainty at the level of atomic contributions with the global error.

\subsection{Atomistic datasets}

We apply this combination of a SOAP-BPNN model and DPOSE architecture to benchmark our approach on four  data sets from the literature. We construct DPOSE ensembles with 64 implicit committee members, and compare them with ensembles of independently trained models with 10 committee members for materials and 5 committee members for the QM9 molecular dataset. 

\paragraph*{Liquid water} We train MLIPs  for liquid water based on the dataset reported in Ref.~\citenum{chengInitioThermodynamicsLiquid2019a},
containing 1593 water structures sampled at different densities and temperatures, with energy and forces computed with the revPBE0 functional \cite{adamoReliableDensityFunctional1999,zhangCommentGeneralizedGradient1998,goerigkThoroughBenchmarkDensity2011} with D3 dispersion corrections \cite{grimmeConsistentAccurateInitio2010, goerigkThoroughBenchmarkDensity2011}. 
Atom-centered descriptors are built using 16 radial functions for the 2-body features, and 5 radial functions and up to 5 angular momentum channels for the power spectrum, using a radial cutoff of 5 \AA.
The atom-centered MLPs -- one for H and one for O atoms -- consist of 2 hidden layers and 64 nodes per layer, and uses SiLU activation functions. We remove the average atomic contributions of the energies prior to training.
Models are trained with minibatch gradient descent using the Adam optimizer \cite{kingmaAdamMethodStochastic2017} in the AMSGrad variant \cite{reddiConvergenceAdam2019b} and batchsize 4, with an initial learning rate of 0.01. We keep the loss weight $\lambda$ fixed at 0.999.
We decay the learning rate with a factor of 0.5 if the force validation loss has not decreased in the last 15 epochs, to a minimal learning rate of 1e-06. We stop the model training if the model has been trained for a maximum of 500 epochs. To stabilizer the optimizer, we use gradient clipping with clipping values of 0.5.
\paragraph*{Lithium thiophosphate} We train MLIPs for the solid-state electrolyte lithium thiophosphate (\ce{Li3PS4}, also abbreviated LiPS or LPS), based on the data set from Ref.~\citenum{gigliMechanismChargeTransport2024a}, that contains 2428 structures of the $\alpha$, $\beta$, $\gamma$ phases of LiPS as well as amorphous structures, computed at the PBEsol level of theory~\cite{perdewGeneralizedGradientApproximation1996,perdewRestoringDensityGradientExpansion2008}. We keep the same MLP and training hyperparameters of the training on liquid water structures. We use adapted powerspectrum hyperparameters from Ref.~\citenum{gigliMechanismChargeTransport2024a} decreasing the radial functions and angular channels to 5.
\paragraph*{Barium titanate} We train a MLIP for BaTiO$_3$, based on the data set from Ref.~\citenum{gigl+22npjcm}, that contains 1458 structures from the three ferroelectric and one paraelectric phases of barium titanate, computed with PBEsol~\cite{perdewGeneralizedGradientApproximation1996,perdewRestoringDensityGradientExpansion2008}.  We keep the same training parameters and MLP hyperparameters as for the training on liquid water structures, but increase the patience of the learning rate scheduler to 20 epochs. We use adapted powerspectrum hyperparameters from Ref.~\citenum{gigl+22npjcm}, decreasing the number of radial functions and angular channels of the power spectrum descriptor to 6.

\paragraph*{QM9 dataset} We train a model of the static formation energy $U_0$ of equilibrium molecular structures from the QM9 data set~\cite{rama+14sd}, containing about 130 thousand local minima configurations, with properties computed at the B3LYP/6-31G(2df,p) level of theory \cite{beckeDensityfunctionalExchangeenergyApproximation1988, beckeDensityFunctionalThermochemistry1993, frischSelfConsistentMolecular1984, leeDevelopmentColleSalvettiCorrelationenergy1988}. 
We use feature hyperparameters adapted from Ref~\citenum{j.willattFeatureOptimizationAtomistic2018a}, with a radial cutoff of 5\,\AA\,and 9 radial functions and angular channels for the powerspectrum. 
We use MLPs with 3 hidden layers and 256 neurons per layer using SiLU activation functions. We restrict our experiments to a random selection of 20000 structures of the QM9 dataset, and train using a batchsize of 32, an initial learning rate of 1e-3, a scheduler patience of 20 epochs and decay factor of 0.5, reducing the initial learning rate to a minimal learning rate of 1e-5. 
Note that the accuracy of this simple model is lower than that of state-of-the-art models \cite{bata+22nips,bigi2023wigner,simeonTensorNetCartesianTensor2024},  yet it is well below chemical accuracy, and a valuable demonstration of the use of a calibrated shallow ensemble for a molecular dataset. 

\begin{table}[btp]
    \centering
    \resizebox{\columnwidth}{!}{%
    \begin{tabular}{clccccc}
    \toprule    
    Dataset  & Model &  MAE  &  RMSE  & NLL $\downarrow$ & RLL $\uparrow$ & MA  $\downarrow$ \\ 
    \midrule
    \multirow{2}{*}{Water}  
       &  DPOSE(NLL)  & 218 & 413 &  -0.40 & 59.0 & 0.04\\
       &  MSE  & 231  & 369 & 0.00   & 33.1 & 0.09 \\
    \addlinespace
    \multirow{2}{*}{\ce{Li3PS4}}  
       &  DPOSE(NLL)  & 86 & 213 &  -1.34 &  63.7  & 0.02  \\
       &  MSE & 92 & 178 & -1.12 &   58.1  & 0.01 \\
    \addlinespace    
    \multirow{3}{*}{BaTiO$_3$}  
       &  DPOSE(NLL) & 10 & 21 & -2.34  &  -34.3 & 0.29 \\
       &  DPOSE(CRPS)& 7 & 12 & -3.41  &  36.7 & 0.05 \\
       &  MSE & 7 & 10 & -3.49 & 31.8 & 0.06  \\
    \addlinespace
    \multirow{3}{*}{QM9(U$_0$)}  
       &  DPOSE(NLL) & 26 & 53 & -1.55  & 2.0 & 0.18\\
       &  DPOSE(CRPS)  & 26 & 51 & -2.08  & 41.4 & 0.04 \\
       &  MSE  & 33 & 54 & -1.63  & 13.2 & 0.09\\
    \bottomrule
    \end{tabular}
    }
    \caption{Accuracy of ensemble models built from independently-trained (MSE loss) members, and of DPOSE models (NLL or CRPS loss, as indicated) for four atomistic datasets. 
    For each model we report test-set MAE and RMSE (in meV, computed for the total energy of structures), and the UQ accuracy as measured by the RLL, the NLL and the miscalibration area.  
    }
    \label{tab:atomistic-benchmarks}
\end{table}

\subsection{Benchmarks}

We begin by comparing uncertainty estimates from committees of separately-trained models (MSE ensemble, a strategy often used in the atomistic modeling literature\cite{schranAutomatedFittingNeural2020, smithLessMoreSampling2018, luUncertaintyEstimatesEquivariantNeuralNetworkEnsembles2023d}) with a DPOSE architecture. 
We don't consider models that directly predict an uncertainty, such as mean-variance estimators, because they do not provide a consistent way of defining errors for the full structure when using an atom-centered decomposition, as we shall discuss further in the next subsection.

\begin{figure}[btp]
    \centering\includegraphics[width=1.0\linewidth]{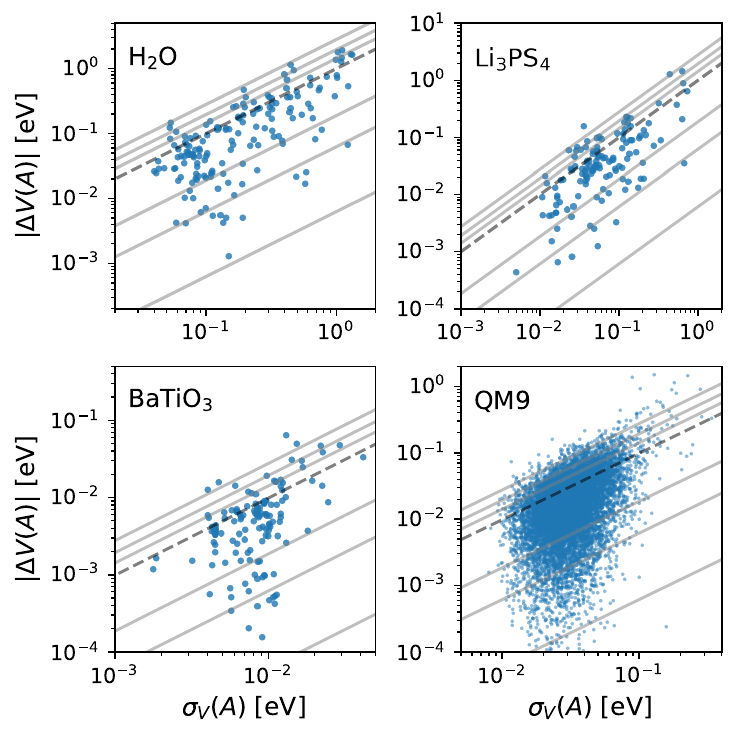} \\ 
    \caption{Predicted-empirical error scatter plots for four atomistic datasets, computed using a DPOSE architecture optimizing using a NLL (\ce{H2O}, \ce{Li3PS4}) and CRPS (\ce{BaTiO3}, QM9) loss. \rev{The gray lines indicate, bottom to top, the 0.5\%, 5\%, 15\%, 85\%, 95\%, 99.5\% quantile lines (and the dashed line the mode) for an ideal Gaussian distribution of the empirical errors, see Fig.~\ref{fig:toy}.}
    All datasets show good correlation between predicted and empirical errors, \rev{with empirical errors consistent with the Gaussian quantiles throughout the predicted uncertainty range.} } 
    \label{fig:atomistic-scatterplots}
\end{figure}

\begin{figure*}[btp]
    \centering
    \includegraphics[width=0.8\linewidth]{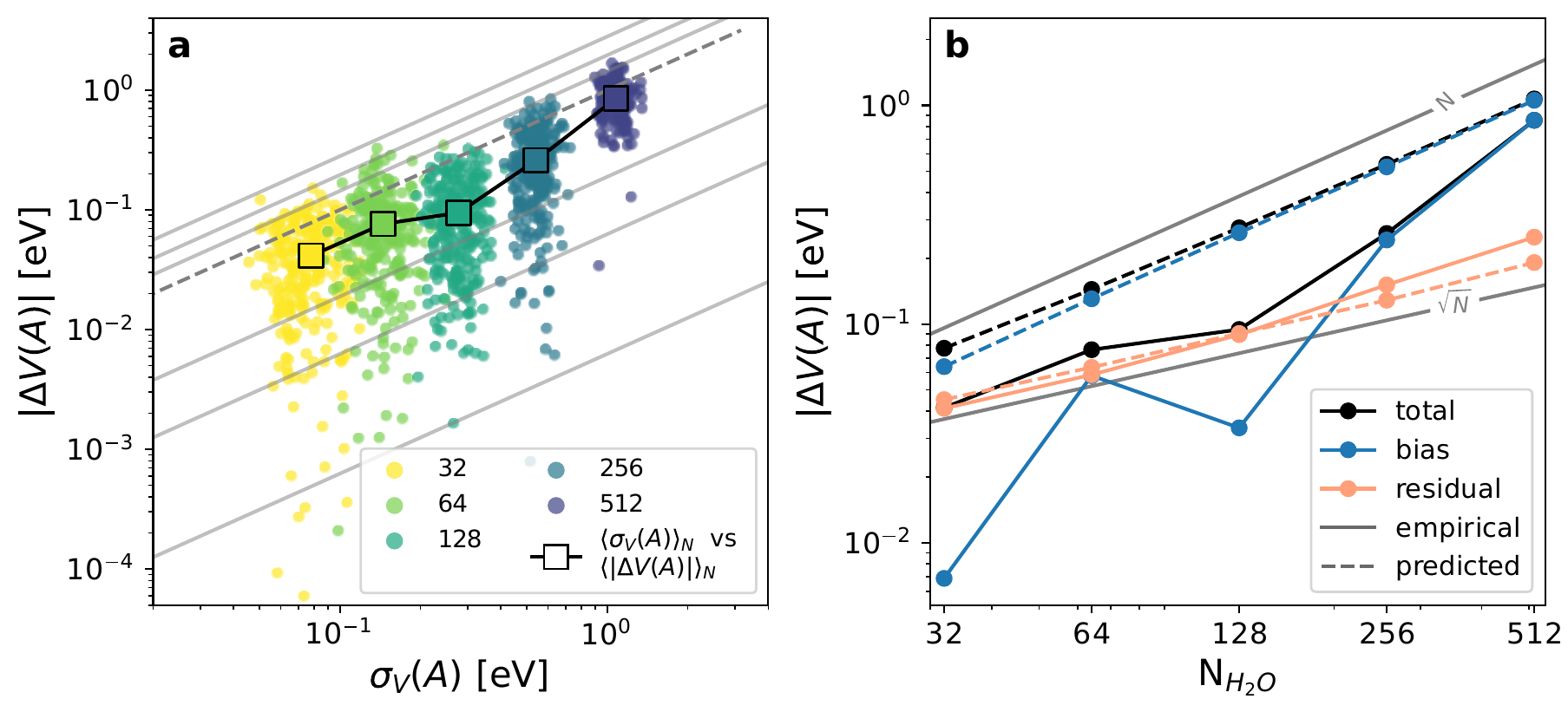}
    \caption{
    (a) Predicted-empirical error scatter plots for constant-density configurations of liquid water simulated at $T=500$~K, using supercells containing different numbers $N$ of molecules. Empty squares correspond to the average errors computed separately for each set of structures.  \rev{The gray lines indicate, bottom to top, the 0.5\%, 5\%, 15\%, 85\%, 95\%, 99.5\% quantile lines (and the dashed line the mode) for an ideal Gaussian distribution of the empirical errors, see Fig.~\ref{fig:toy}.} (b) Scaling with system size of total, bias, and residual errors, including both predicted and empirical values. Errors are computed as follows. 
    Total empirical error: $\langle|\bar{V}-V_\text{ref}|\rangle_N$; empirical bias: $|\langle \bar{V}-V_\text{ref}\rangle_N|$; empirical residual: $\langle|\bar{V}-V_\text{ref}- \langle \bar{V}-V_\text{ref} \rangle_N |\rangle_N$; total predicted error: $\langle\operatorname{std}_k (V^{(k)}-\bar{V})\rangle_N$; predicted bias: $\operatorname{std}_k(\langle V^{(k)} - \langle \bar{V}\rangle_N\rangle_N)$; predicted residual: $\langle \operatorname{std}_k(V^{(k)}-\langle V^{(k)}\rangle_N) \rangle_N$. $\operatorname{std}_k$ indicates the standard deviation over the members of the ensemble, and $\langle\cdot\rangle_N$ averaging over all structures with $N$ molecules. 
    }
    \label{fig:size-scaling}
\end{figure*}

We observe that (1) calibration is essential to obtain meaningful error estimates with an ensemble of MSE-trained models (non-calibrated MSE models show very large, negative RLL values, similar to what seen in Table~\ref{tab:general-benchmarks-RLL}) and (2) using a NLL loss can lead to  degradation of the energy accuracy compared to training against a MSE loss. 
This is a well-known problem, due to the fact that the NLL loss is less sensitive to outlier structures with large prediction errors, as long as they are associated with a large uncertainty, leading to local minima in the loss that have both large prediction error and high NLL. 
A possible strategy to mitigate this issue is to use a different metric in the loss, such as the $\beta-$NLL \cite{seitzerPitfallsHeteroscedasticUncertainty2021}. Following Ref.~\citenum{gneitingCalibratedProbabilisticForecasting2005}, we use the continuously ranked probability score (CRPS)\cite{brownAdmissibleScoringSystems1974, hersbachDecompositionContinuousRanked2000a}, that can be computed analytically for a  normal distribution \cite{gneitingCalibratedProbabilisticForecasting2005}, and reads
\begin{equation}
\begin{split}
    &\rm{CRPS}_{\mathcal{N}}(\Delta y, \sigma) = \\
    &\sigma \Big\{ \frac{\Delta y}{\sigma}\Big[2\Phi\Big(\frac{\Delta y}{\sigma}\Big) -1 \Big] + 2\varphi\Big(\frac{\Delta y}{\sigma}\Big) - \frac{1}{\sqrt{\pi}}\Big\}, \\
 \end{split}
\end{equation}
where $\varphi$ and $\Phi$ are the normal probability and cumulative  distribution function, with zero mean and unit variance, calculated at ${\Delta y}/{\sigma}$.
In the SI we demonstrate on a sinusoidal toy system that training on the NLL loss can lead to poor convergence~\cite{seitzerPitfallsHeteroscedasticUncertainty2021} and employing the CRPS loss mitigates the risk of converging to stable local minima with poor performance metrics. 
As shown in Table~\ref{tab:atomistic-benchmarks}, training on a CRPS loss improves the models RMSE and MAE, and even the validation NLL values, giving a clear indication that the NLL loss is susceptible to the presence of local minima.
When using the better-behaved CRPS loss,  DPOSE provides a consistently better uncertainty model than a calibrated ensemble of indipendently models trained with an MSE loss, in terms of all the diagnostics we considered. 
The predicted-empirical error scatter plots in Fig.~\ref{fig:atomistic-scatterplots} are also indicative of well-calibrated, informative uncertainty models, \rev{with uncertainty predictions spanning a range of at least one order of magnitude and emprirical errors that are consistent with a normal distribution.} 
The \ce{BaTiO3} model has a narrower predicted uncertainty range, which is consistent with the training set being composed of a homogeneous collection of thermally distorted crystalline configurations, with a less pronounced heteroskedactic behavior than the more diverse datasets used in the other benchmarks. 

\subsection{Size extensivity and forces}

The atom-centered decomposition~\eqref{eq:atom-centered-v} has important implications when it comes to defining errors for the total energy $V(A)$ that is built combining the atomic energy for the environments, $v(A_i)$. 
This is highly non-trivial for uncertainty models that explicitly generate a variance defined at the level of the environments, $\sigma_v(A_i)$: both the total potential $V(A)$ and its derivatives are the sum of many terms, and the corresponding errors cannot be computed without knowing the correlations between individual contributions. 
Even though many papers discuss the closely-related difficulty in deriving a consistent definition of force errors \cite{buskGraphNeuralNetwork2023a,carreteDeepEnsemblesVs2023, jorgensenCoherentEnergyForce2023}, based on an atom centered model of energies and their uncertainties, the relevance of this problem is not fully appreciated and is worth investigating in more detail.
Consider two limiting cases: one is a crystalline structure in which all atomic sites are equivalent because of symmetry; the other a dilute gas of molecules, each with a randomly chosen composition and/or structural deformation. 
In the former case, a symmetry-adapted ML model will predict the exact same atomic energy for each site, and so the errors for a crystal containing $N$ atoms should be just $\sigma_V=N\sigma_v$. 
In the second case, errors should be uncorrelated and combine in quadrature, so $\sigma^2_V(A)\approx\sum_i \sigma^2_v(A_i)$, and asymptotically $\sigma_V\propto \sqrt{N}\sigma_v$. 

\begin{figure*}[pbt]
    \centering
    \includegraphics[width=0.85\linewidth]{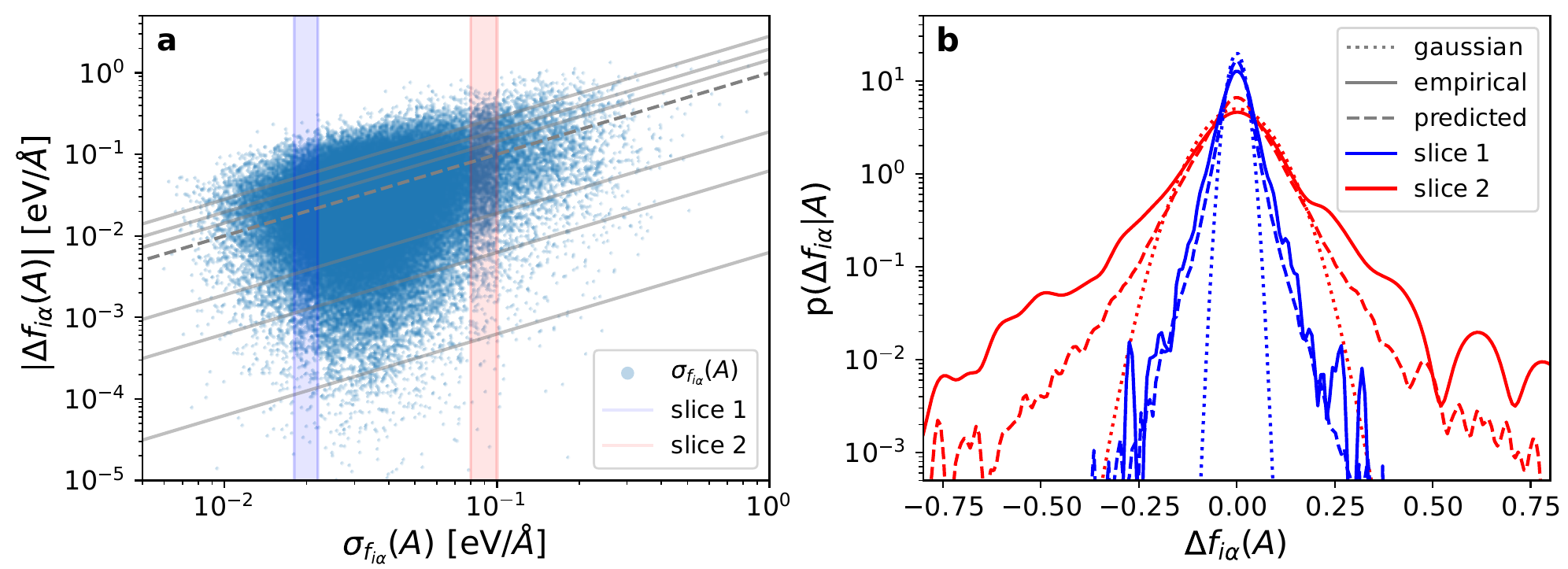}
    \caption{(a) Predicted-empirical error scatter plot for the components of the force acting on the atoms of a water simulation. \rev{The gray lines indicate, bottom to top, the 0.5\%, 5\%, 15\%, 85\%, 95\%, 99.5\% quantile lines (and the dashed line the mode) for an ideal Gaussian distribution of the empirical errors, see Fig.~\ref{fig:toy}.} The two shaded areas indicate regions used to select the points used in the bottom panel.
    (b) Histograms of the empirical errors $\Delta f_{i\alpha}$ (full lines) and of the distribution of ensemble members around the mean prediction, corresponding to the predicted $p(f_{i\alpha}|A)$ (dashed lines). The dotted lines correspond to a Gaussian distribution with the equivalent variance.  }
    \label{fig:water-force}
\end{figure*}

In realistic cases, the behavior may be somewhere in between these extremes: atomic environments within a structure may share some characteristics, and be affected by the same type of prediction errors, while having random distortions that lead to uncorrelated error contributions. 
Fortunately, the use of a direct ensemble propagation framework means that the existence of correlations between the model predictions is automatically accounted for. The expression~\eqref{eq:committee-total-v} for $\sigma_V^2(A)$ relies on the individual atomic predictions, rather than on a hard-to-define propagation of the atomic energy uncertainties. 
To demonstrate this, we construct a collection of structures of liquid water, using supercells of different size and a constant number density. 
We perform long molecular dynamics simulations at constant volume and temperature equal to 500K, and collect uncorrelated structures that we re-compute using electronic-structure parameters consistent with those used in the generation of the train set. 
The scatter plot for the predicted and empirical errors resolved over structure size shows a good correlation (Fig.~\ref{fig:size-scaling}a) even though the predictions are miscalibrated (underconfident). 
The error observed on structures of varying size can be decomposed into two contributions. 
Given that the MLIP is trained on a heterogeneous dataset with different temperatures and densities, and that all structures used in this exercise have the same molar volume, it is reasonable to expect that there will be a constant \emph{bias} error, associated with the prediction of the mean energy at this fixed density. 
This bias can be estimated by taking the average observed error over each set of structures, a contribution to the overall errors that scales roughly as $N$ (Fig.~\ref{fig:size-scaling}b).
\rev{The empirical bias is just the absolute difference between the predicted and reference potential over all structures of a given size $N$, $|\langle \bar{V}\rangle_N - \langle V_\text{ref}\rangle_N|$.}
By subtracting the bias from each configuration, one is left with a term that reflects the energy fluctuations within each snapshot, that are a consequence of the random distortion of each molecule. 
The errors on this residual term, therefore, scale as $\sqrt{N}$.
This experiment demonstrates that, in this realistic example, the relationship between local and total uncertainties cannot be captured by a single scaling law -- which would make it difficult to define an accurate estimate for $\sigma_V$ based on predictions of the atomic energy uncertainties $\sigma_v$

The error predicted through Eq.~\eqref{eq:committee-total-v}naturally captures these two components, that can be computed separately by simply repeating the manipulations we performed to separate the bias and the residual in the reference values on individual ensemble predictions. 
\rev{ 
For example, the predicted bias can be estimated by applying direct propagation to this ensemble average, i.e. estimating the mean $\langle V^{(k)}\rangle_N$ over all structures of size $N$ for each ensemble member separately. }
The mean energy over multiple frames is affected by a large error: this is the correlated term that will also be shared by all environments across one structure, and exhibit a linear scaling with system size. 
The error on the residual part, instead, is largely uncorrelated between different atom-centered environments, and scales roughly as $\sqrt{N}$. 
The bias error is overestimated, which leads to  underconfident error estimates, and to the fact that the overall predicted error exhibits a leading-order linear scaling already at the smaller system sizes. 
\rev{This effect can be also made apparent by showing parity plots of the predicted and reference potentials of the individual committee members (see SI).}

Similar arguments apply to the calculation of force errors, that are not directly related to the error on the energy of a given configuration, but depend on the correlation between the errors on configurations that are slightly deformed with respect to one another. 
As shown in Figure~\ref{fig:water-force}a, the DPOSE estimators~\eqref{eq:committee-total-f} are capable of detecting force components that are affected by a large error. 
The estimates are also well-calibrated, despite the fact that no explicit force uncertainty term is included in the training loss~\eqref{eq:atom-centered-loss}, and the fact that for these structures the energy estimator is underconfident. 
The near-constant error that is common to all structures in this dataset, associated with the mean density of the liquid, has very little impact on the forces, that are dominated by short-range interactions, and that are largely uncorrelated between different atoms in the liquid. 
This example also allows us to comment on an important advantage of an ensemble approach.
The distribution of empirical force errors (computed using~\eqref{eq:ensemble-nonpara}) is strongly non-Gaussian, and has a large tail that is clearly visible in Fig.~\ref{fig:water-force}a, where many samples are far above the 99\%{} quantile line even though the mode of the distribution appears well calibrated.
The DPOSE architecture \rev{allows one to compute a non-parametric ensemble distribution through Eq.~\eqref{eq:ensemble-nonpara}} that is in good  qualitative agreement with the non-Gaussian nature of the empirical error distribution (Figure~\ref{fig:water-force}b). Thanks to the direct propagation of the ensemble members, the effect of this non-Gaussian distribution on any derived quantity would also be properly accounted for.

\subsection{Uncertainty propagation}

The calculation of errors on total energies and forces, discussed in the previous section, can be regarded as a successful application of uncertainty propagation, as the ``raw'' ensemble  of atomic contributions is used to determine the error on $V(A)$ and $f_{i\alpha}(A)$.
Ensemble propagation can also be applied to more complicated cases -- for instance to the determination of average structural and thermodynamic properties, as shown in Ref.~\citenum{imba+21jcp} for the case of MSE ensembles without weight sharing. 
We use the shallow ensemble potential for liquid water discussed in the previous section to run for 300ps  a molecular dynamics simulation of 256 molecules of liquid water at constant volume and  $T=300$K, performed using the i-PI package.\cite{kapi+19cpc} 

\begin{figure}[bp]
    \centering
    \includegraphics[width=\linewidth]{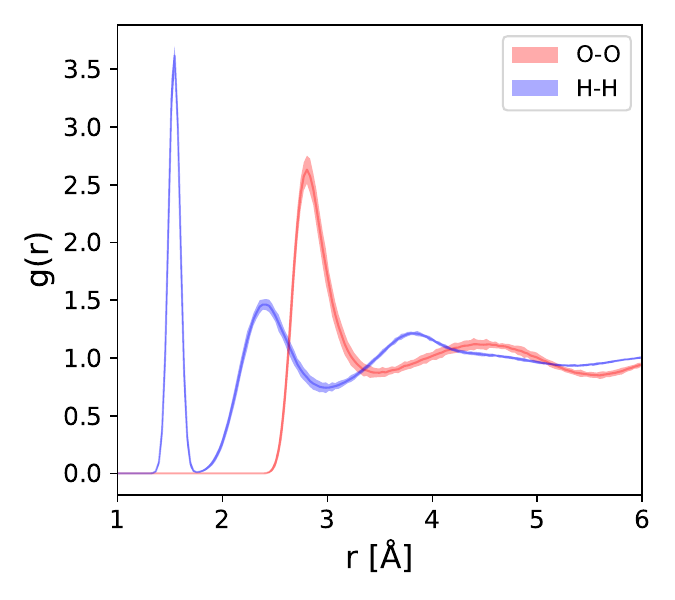}
    \caption{O-O and H-H pair correlation functions for a 256-molecule simulation of liquid water at 300~K. The thin line and the shading indicate the mean and confidence intervals obtained from ensemble propagations using the cumulant expansion expression~\eqref{eq:reweight-cea}.
    }
    \label{fig:enter-label}
\end{figure}

We then evaluate the \emph{pair correlation function} $g(r)$, corresponding to a suitably normalized histogram of the interatomic distances $r$, that reports on mean structural features and can be measured experimentally by x-ray or neutron diffraction\cite{soperRadialDistributionFunctions2000, skinnerStructureWaterCompressibility2014, okhulkovRayScatteringLiquid1994}. 
The histogram is computed over snapshots that are distributed according to $e^{-\bar{V}/k_\text{B}T}$, but one can estimate the histograms that would have been generated by each committee member by determining, for each configuration, a weight computed following the linearized reweighting expression~\eqref{eq:reweight-cea}. 
Importantly, this means that the errors in the $g(r)$ are not constant, but different correlations and different length scales exhibit different confidence intervals, reflecting the accuracy of the interatomic potential in reproducing different types of interactions. 
For example, inter-molecular interactions, that are responsible for the O-O correlations and the large-$r$ parts of the O-H and H-H correlations, show larger errors than the intra-molecular, short-$r$ parts -- which is consistent with the fact that the potential has large errors for density fluctuations, and smaller errors in the short-range-dominated force components.

Linearizing the reweighting expression is essential to obtain statistically-efficient averages, particularly at large system sizes: the statistical error in a non-linearized reweighting approach grows exponentially with the mean-square discrepancy between potentials, that grows linearly with system size\cite{ceri+12prsa}.
When using Eq.~\eqref{eq:reweight-cea}, instead, the statistical error does not increase with system size, and one can estimate that, as expected, the ML error on this relatively large box is comparable to that computed for a smaller box (see SI). 

The cumulant expansion approximation should be used with some care, as we demonstrate in the case of the constant-volume heat capacity. Note in passing that this is a quantity that cannot be assessed accurately without a quantum mechanical description of the hydrogen nuclei\cite{vega+14jcp}, and so we only discuss it to show the limits of Eq.~\eqref{eq:reweight-cea}.
In a classical molecular dynamics simulation, the heat capacity can be computed from the fluctuations of the potential energy, 
\begin{equation}
C_V=\frac{\left<(V-\left<V\right>)^2\right>}{k_\text{B}T^2} + \frac{3}{2} N k_\text{B}.
\end{equation}
Using the same simulation of 256 water molecules at 300~K, we find an estimate of the heat capacity per water molecule of $14.51 k_\text{B}$. 
Using Eq.~\eqref{eq:reweight-cea} to evaluate the error on $C_V$ is problematic because -- even in cases in which $V^{(k)}$ and $(V^{(k)}-\bar{V})$ are distributed as a correlated Gaussian, which is the condition underlying the approximation -- $(V^{(k)}-\left<V^{(k)}\right>)^2$ is strongly non-Gaussian (see the SI for a figure demonstrating this behavior): indeed, naive application of Eq.~\eqref{eq:reweight-cea} yields an estimate of   $C_V/k_\text{B}=-0.8\pm 20.9$, which is manifestly absurd: a heat capacity cannot be negative, and nothing in the behavior of the committee members suggests that the error on fluctuations should be exceptionally large. 
To verify that this is indeed a consequence of the incorrect application of the cumulant expansion expression, we can use a generalization of it, that provides an analytical estimate of all re-weighted moments of a random variable $y$. 
Assuming only Gaussian behavior for $y$ and the logarithm of the weighting factor, one can obtain (see Ref.~\citenum{ceri+12prsa} for an outline of the derivation)
\begin{equation}
\label{eq:reweight-cea-moment2}
\left<(y-\langle y \rangle)^n\right>_{V^{(k)}} \approx 
\left<(y-\langle y \rangle)^2\right>_{\bar{V}}^{\frac{n}{2}} 
\frac{2^n\Gamma(\frac{n+1}{2}) (1+(-1)^n)}{2\sqrt{\pi}}
,
\end{equation}
For the second centered moment of the ensemble potentials, one gets simply
\begin{equation}
\label{eq:reweight-vk2}
\left<(V^{(k)}-\langle V^{(k)} \rangle_{V^{(k)}})^2\right>_{V^{(k)}} \approx 
\left<(V^{(k)}-\langle V^{(k)} \rangle_{\bar{V}})^2\right>_{\bar{V}},
\end{equation}
i.e. one should look at the spread in the fluctuations of individual members of the potential, for configurations collected along the trajectory driven by $\bar{V}$ and without adjusting for the distortion of the distribution. 
With this corrected expression, we obtain $C_V/k_\text{B}= 14.64\pm 0.47$: the propagated ensemble mean value is consistent with the heat capacity computed from the mean potential, the propagated uncertainty is not overestimated dramatically. 
This error is comparable to that associated to statistical convergence of the average values, emphasizing the importance of incorporating \emph{all} sources of error when assessing the overall reliability of the output of molecular simulations. 

\begin{figure}[tbp]
    \centering
    \begin{tabular}{cc}
        \includegraphics[width=\linewidth]{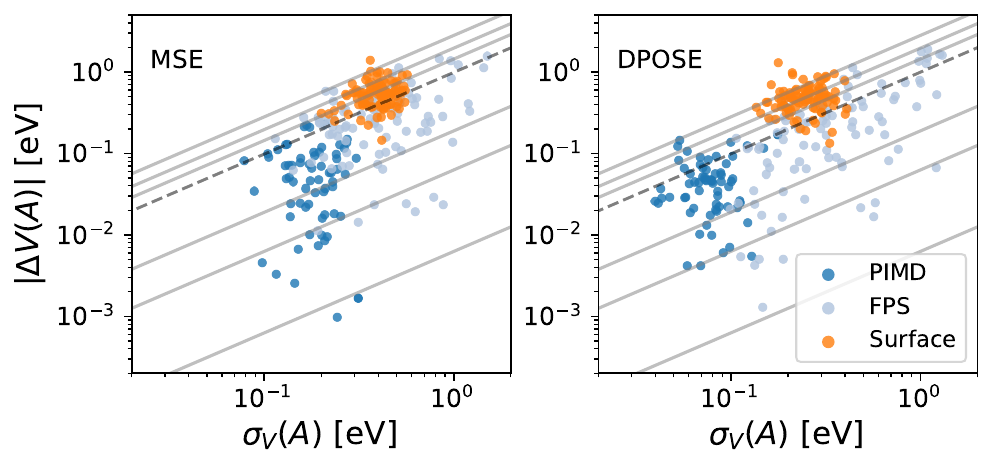}
    \end{tabular}
    \caption{Predicted-empirical error parity plots for liquid water configurations, obtained with a MSE ensemble (left) and with DPOSE (right).
    Different colors indicate subsets of the validation structures extracted from molecular dynamics sampling (PIMD, dark blue), from farthest point sampling (FPS, light blue) and from surface configurations (orange). 
    \rev{The gray lines indicate, bottom to top, the 0.5\%, 5\%, 15\%, 85\%, 95\%, 99.5\% quantile lines for an ideal Gaussian distribution of the empirical errors, see Fig.~\ref{fig:toy}.}
    }
    \label{fig:surfaces}
\end{figure}

\subsection{Extrapolative predictions}

As recently highlighted in the work of Lu and coworkers \cite{luUncertaintyEstimatesEquivariantNeuralNetworkEnsembles2023d},
out of-distribution-uncertainty quantification remains a challenging problem in applications to atomistic modeling.
To assess the behavior of a DPOSE model in the case of extrapolative predictions, we compare it with that of a calibrated ensemble of models independently trained based on the MSE. 
We use as demonstrative example a set of 100 liquid water surface structures, obtained from molecular dynamics simulations with the shallow ensemble potential.
To interpret the predicted-empirical error scatter plots (Fig.~\ref{fig:surfaces}) it is important to recall that the training data for this model comprises room-temperature MD simulations as well as highly-distorted structures obtained by farthest point sampling (FPS), that include self-dissociated molecules as well as large density fluctuations and voids. 
The predicted (and observed!) errors for these surface structures, that differ qualitatively from those found in the training set, are comparable to those observed for the highly-distorted FPS structures, and much larger than for those sampled from room-temperature simulations. 
Thus, both models succeed in capturing the qualitative observation that these surface structures -- despite being obtained from room-temperature simulations -- have errors that are closer to those seen for distorted configurations, even though the true magnitude of the error is significantly underestimated. 
The high degree of diversity of the FPS structures underpins the remarkable stability of the model, that is capable of running simulations with slab geometries despite being trained exclusively on bulk configurations. 

\begin{figure}[tbp]
    \centering
    \includegraphics[width=\linewidth]{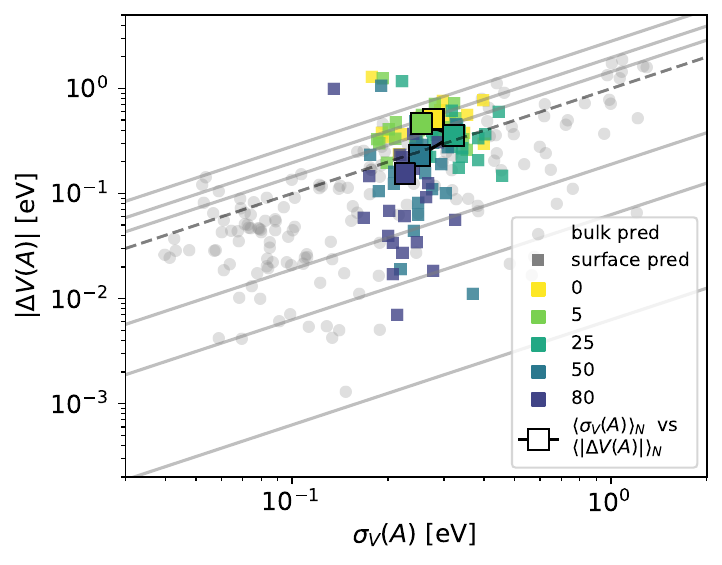}
    \caption{Evolution of the error distribution for 20 liquid/vapor configurations, computed with a DPOSE model as a growing number of this kind of structures are included in the training set. The grey circles are the results the bulk structures shown in Fig.~\ref{fig:surfaces}, and squares are color-coded to indicate the different stages of training set augmentation. Larger squares correspond to the mean over these validation structures, and provide a clearer indication of the trends. }
    \label{fig:surface-active}
\end{figure}

Comparing DPOSE with the MSE ensemble, one can see a higher discriminating power, with the errors of the surface structures being clearly separated from those of the low-temperature PIMD configurations. 
It would therefore be possible to use the predicted uncertainties to identify high-error structures, and to implement an active-learning loop. 
\rev{To demonstrate how including additional structures affects the predicted and empirical errors, we repeated training and validation exercises for the water model, including an increasing fraction of the structures that include a surface. 
As shown in Fig.~\ref{fig:surface-active}, after adding 25 of these configurations both the accuracy of the model and its ability to make well-calibrated uncertainty estimates improve substantially. Even though the empirical errors remain much higher than those for the bulk configurations -- probably also due to the difficulty in predicting surface energies of a polar liquid using a ML model that is limited to short-range interactions\cite{nibl+21jcp,zhai+23jcp} -- DPOSE becomes able to estimate them accurately, which is important to improve the active-learning loop, and ultimately to ensure the overall reliability of the ML protocol. 
}

\begin{figure}[btp]
    \centering
    \includegraphics[width=\linewidth]{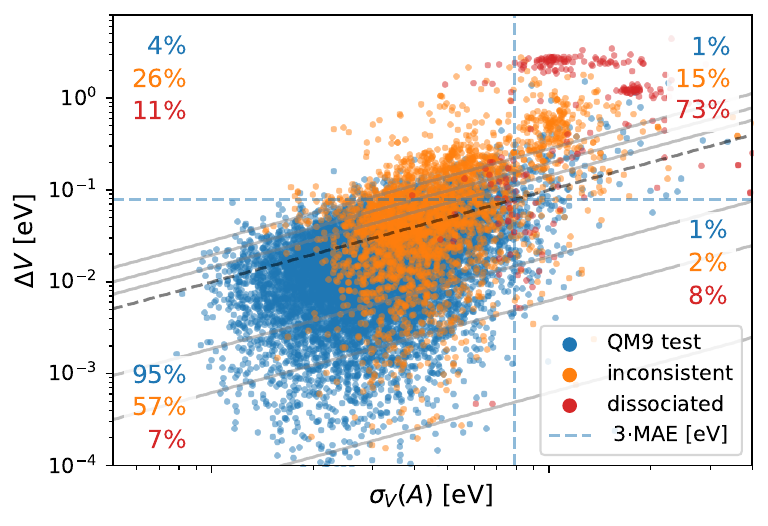}
    \caption{\rev{Predicted-empirical error parity plot for the QM9 dataset. The blue points are the same in-set test structures shown in Fig.~\ref{fig:atomistic-scatterplots}.
    The orange points are the 3'000 structures that fails a bonding ``consistency check'', and the red points the subset of those in which the target molecule breaks down into several fragments. 
    The dashed lines corresponds to a an error of three times the test MAE, that is taken as a threshold to propose candidates for active learning. The lines divide space in four quadrants, that in a classification language can be interpreted as true positives (upper right), true negatives (lower left), false positives (lower right) and false negatives (upper left). The percentages on the graph indicates the fractions of each subsets of structures that fall within each quadrant.  
    }
    }
    \label{fig:qm9-active}
\end{figure}

\rev{
As a further demonstration of the accuracy of DPOSE error estimation in an extrapolative regime, we consider the case of the QM9 ``inconsistent'' structures. 
The QM9 dataset contains about 3'000 molecules that, following geometry optimization, display bonding patterns that are inconsistent with the SMILES string they were built from.\cite{rama+14sd} 
Some of these dissociate into fragments (typically releasing stable molecules such as nitrogen), while others undergo major chemical rearrangements while remaining intact.  
As shown in Fig.~\ref{fig:qm9-active}, the DPOSE model is generally overconfident for these extrapolative predictions, although it correctly estimates these inconsistent structures as having larger errors than the typical QM9 structure. 
One way of assessing how useful the extrapolative performance of the DPOSE model are is to stipulate that structures with an error that is larger than three times the test-set MAE should be tagged for further training, and checking how the classification performed based on the predicted error compares with that based on the empirical errors. 
For the internal test set, the vast majority of structures fall below this threshold, and 96\%{} of structures are classified correctly. 
This percentage falls to 72\%{} when considering the inconsistent structures, with 26\%{} of ``false negatives'' that are especially problematic -- as they indicate that the model is overconfident in its extrapolative capabilities. 
It is however reassuring that for the more extreme case of dissociated structures 80\%{} of the test configurations are classified correctly, and the rate of false negatives decreases to 11\%{}. 
This experiment shows that shallow ensembles can deliver useful levels of uncertainty estimation also in the far extrapolative regime, even though the reliability of the estimates is lower than for the in-set case, and it might be necessary to use an active-learning strategy to improve the error estimates, as well as the actual prediction accuracy. 
}

\section{Conclusions}

Supplementing the predictions of a machine-learning model with uncertainty estimates is extremely important to increase the confidence with which it can be used in applications. 
We introduce a framework to do so based on ensembles of models sharing all but the last-layer weights, that we show to be inexpensive, accurate and that -- crucially -- provides a simple and reliable strategy to propagate uncertainty on derived quantities. 
While we justify this last-layer approximation empirically, it is worth mentioning that a recent work, that offers a probabilistic interpretation of uncertainty estimates based on the last layer approximation,\cite{bigi2023pr} provides a theoretical support to this choice. 
We confirm the importance of calibrating the error predictions, either by explicitly training on a loss that incorporates the uncertainty of training structures, or by a post-hoc scaling factor correcting the spread of the ensemble around their mean. 
While we find that using the NLL as a loss does not degrade the raw accuracy of the model predictions for a set of benchmark datasets, in some of our atomistic experiments a NLL loss increases by up to a factor of two the RMSE. 
In these cases, we find that using an alternative metric (the CRPS) as a loss improves the model accuracy and even improves the NLL. This suggests that poor convergence of the iterative optimization of the NLL loss (already observed in several prior works \cite{skafteReliableTrainingEstimationb, seitzerPitfallsHeteroscedasticUncertainty2021, takahashiStudenttVariationalAutoencoder2018, stirnVariationalVarianceSimple2020a}) is to blame for the degraded performance. 
Using an approximate Hessian of the loss for a pre-trained model to sample the weights of a shallow calibrated ensemble  would be one possible strategy to circumvent the difficulties of optimizing an NLL-like loss.\cite{bigi2023pr} 

Experiments on atomistic datasets allow us to showcase a distinctive feature of the DPOSE framework: the ease of propagating uncertainty through simple - as well as highly non-trivial - post-processing steps. 
This is essential for models that are built as additive combinations of ``local'' contribution to a total quantity such as the energy. 
Uncertainty estimators based on an explicit variance prediction do not offer a straightforward recipe to combine local errors into a global uncertainty estimate. 
Calibrated ensembles, instead, capture transparently the interplay between a systematic, size-extensive bias and a stochastic ``thermodynamic'' contribution, that scale differently with system size, and make it easy to compute errors on gradient quantities (e.g. forces acting on the atoms). 
We also demonstrate that a shallow ensemble architecture is effective when computing average quantities over a molecular dynamics trajectory, similar to what was shown previously for ensembles of independently-trained models\cite{imba+21jcp}.
\rev{Even though the extrapolative predictions of the DPOSE models are not as accurate as those for a in-set validation, they are sufficiently precise to identify out-of-sample predictions, and to improve the uncertainty estimation accuracy as well as the prediction error, by adding just a few additional configurations to the training set. 
Combined with the low computational cost that allows  performing on-line uncertainty estimation, DPOSE is thus well-suited for active-learning frameworks.} 
Overall, we think that these benchmarks demonstrate that our approach addresses the requirements of ease of implementation, general applicability, low computational overhead and availability of propagated uncertainties, making it ideal for atomic-scale modeling, to drive  active-sampling strategies, and more broadly for any application that requires manipulating in a non-trivial way the outputs of several (possibly-correlated) ML model predictions.

\grayout{
\section{TODOs}
\begin{itemize}
\item \sout{Finalize figure 1, 2, 3, 4, 5, 6, 7, 8}
\item \sout{Finalize tables I, II}
\item \sout{Finish description of datasets}
\item \sout{Finish discussion of figures of merit}
\item Add all references
\end{itemize}

\section{TODOs - possibly after submission}
\begin{itemize}
\item Prepare data record, including all datasets (making sure they are well annotated and reference the original publications)
\item toy example notebook
\item demo models for all datasets. 
\item {\color{blue} Add "failing smiles" part of QM9 }
\item Add RLL for each subset in the water surface plot for the two models
\item In the SI, when discussing convergence with system size, show results for an explicit exponential reweighting, demonstrating the instability of the base expression with system size
\item Merge i-PI PR and prepare QM example for the i-PI paper, 
\end{itemize}
}

\vspace{3mm}
\section*{Supporting data}

The data used for the UCI benchmarks and the atomistic examples is available from the original publications. 
A demonstrative implementation of the DPOSE framework, including the examples discussed here, is available at \url{https://github.com/bananenpampe/DPOSE}.

\begin{acknowledgments}
The authors would like to thank Filippo Bigi, Sanggyu Chong and Federico Grasselli for insightful discussion and comments on an early version of the manuscript, Mariana Rossi and Shubham Sharma for an ongoing collaboration on the implementation of ensemble UQ in i-PI, and Philip Loche for help with DFT calculations. We would like to thank Marcel Langer and Davide Tisi for a careful review of an early version of this manuscript.
Financial support by the Swiss National Science Foundation (Project 200020\_214879) is gratefully acknowledged.
\end{acknowledgments}

\end{document}